\documentclass[aps,pre,twocolumn]{revtex4-1}
\usepackage{graphicx}
\usepackage{dcolumn}
\usepackage{bm}
\usepackage{amsmath}
\usepackage{mathrsfs}
\begin{document}

\title{Robustness of scale-free networks to cascading failures induced by fluctuating loads}

\author{Shogo Mizutaka}
\email{s.mizutaka@eng.hokudai.ac.jp}
\author{Kousuke Yakubo}
\email{yakubo@eng.hokudai.ac.jp} \affiliation{Department of
Applied Physics, Hokkaido University, Sapporo 060-8628, Japan}

\date{\today}
\begin{abstract}
Taking into account the fact that overload failures in
real-world functional networks are usually caused by extreme
values of temporally fluctuating loads that exceed the
allowable range, we study the robustness of scale-free networks
against cascading overload failures induced by fluctuating
loads. In our model, loads are described by random walkers
moving on a network and a node fails when the number of walkers
on the node is beyond the node capacity. Our results obtained
by using the generating function method shows that scale-free
networks are more robust against cascading overload failures
than Erd\H{o}s-R\'enyi random graphs with homogeneous degree
distributions. This conclusion is contrary to that predicted by
previous works which neglect the effect of fluctuations of
loads.
\end{abstract}


\maketitle

\section{INTRODUCTION}
\label{sec:1}
Modern human societies are supported by various functional
networks, such as power grids, the Internet, road systems, and
corporate transaction networks \cite{Krol15,Costa11}. Since the
function of a network is guaranteed by its global connectivity,
a decomposition of the network into disconnected components by
failures on network elements (nodes or links) might induce the
breakdown of the function, which causes a fatal damage to our
daily life. It is thus crucial to elucidate what type of
network topology is resilient to failures. In this context, the
vulnerability of complex networks against random failures and
targeted attacks has been extensively studied
\cite{Cohen00,Cohen01,Newman02,Holme02a,Boccaletti06,Dorogov08}.
In these studies, failed nodes are removed from a network at
the same time. In addition to such simultaneous failures, many
of actual breakdowns are driven by a chain of failures
triggered by initial failures of a single or a few nodes.
Initial failures and subsequent avalanche of failures in
functional networks are often induced by loads exceeding node
capacities. For instance, a bankruptcy of an insolvent company
and subsequent chain bankruptcies due to redistributed debts of
the bankrupt company can be regarded as a process of such
\textit{cascading overload failures} in a corporate transaction
network.

In order to describe how networks lose their global
connectivity by cascading overload failures, Motter and Lai
proposed a model in which the load at a node is given by the
betweenness centrality of the node \cite{Motter02}. Their model
predicts that the initial removal of the highest degree (or
highest load) node leads a large scale cascade and scale-free
networks are more fragile against cascading overload failures
than homogeneous networks. The vulnerability of scale-free
networks has also been found in other models of cascading
failures \cite{Holme02b,Crucitti04,Wu07,Bao09,Xia10,Dou10}. In
these models, overload failures are caused by
\textit{non-fluctuating} loads determined uniquely by the
network structure that exceed the capacity of each node. It is,
however, quite general that the load on a node fluctuates
temporally \cite{Menezes04,Meloni08,Zhou13}. An overload
failure takes place when an instantaneous value of fluctuating
load exceeds the capacity, as in the cases of river flooding or
financial collapse in a country \cite{Embrechts97}. The
property of cascading overload failures induced by
\textit{extreme events} might be largely different from that by
non-fluctuating loads.

However, the robustness of networks against cascading failures
induced by fluctuating loads has not been widely argued so far
\cite{Heide08}. Among several ways to describe fluctuating
loads, Kishore \textit{et al.}~modeled them by random walkers on a
network \cite{Kishore11,Kishore12}. They calculated the
overload probability that the number of random walkers on a
node exceeds the predetermined node capacity. Moreover,
applying this theory, the network robustness against
non-cascading overload failures has been studied, in which
nodes are simultaneously removed once according to the overload
probability \cite{Mizutaka13,Mizutaka14}. Although these
theories take into account the temporal fluctuations of loads,
cascade processes triggered by the initial overload failures
have not been considered. In this paper, we examine the
robustness of complex networks against cascading overload
failures induced by the extreme value of fluctuating loads.
Adopting the random walker model proposed by Kishore \textit{et
al.}~and their theory of the overload probability
\cite{Kishore11,Kishore12}, we present a simple model to
describe cascades of overload failures caused by fluctuating
loads and calculate the size of the giant component after
completing the cascade by using the generating function
formalism. Our main result shows that scale-free networks are
more robust against cascading overload failures than
homogeneous networks, which is contrary to that predicted by
previous works \cite{Motter02,Holme02b,Crucitti04,Wu07,Bao09,
Xia10,Dou10}.

The rest of this paper is organized as follows. In
Sec.~\ref{sec:2}, we present a model to describe cascading
overload failures on a complex network based on the random
walker model proposed by Kishore \textit{et al.}~\cite{Kishore11,
Kishore12}. In Sec.~\ref{sec:3}, we explain the method to
calculate the size of the giant component after completing the
cascade process by utilizing the master equation for the
probability of a node to have the initial degree $k_{0}$ and
the degree $k$ at a cascade step $\tau$ and the generating
function formalism. Our results are presented in Sec.~\ref{sec:4}.
Section \ref{sec:5} is devoted to the summary and concluding
remarks.

\section{MODEL}
\label{sec:2}
In a functional network, some sort of ``flow" is often required
to realize its functionality, and at the same time the flow
plays a role of a ``load" in the network, such as electric
current in a power grid or packet transfer on the Internet. The
relation between the average and fluctuation of flow in such
networks has been investigated empirically and theoretically
\cite{Menezes04,Meloni08,Zhou13}. These studies elucidated that
the flux fluctuations at a node have the same scaling behavior
with the fluctuations of the number of random walkers on the
node. Inspired by this fact, Kishore \textit{et al.}~modeled
fluctuating loads by random walkers moving on a network
\cite{Kishore11,Kishore12}, where the number of walkers
indicates the amount of loads. The stationary probability to
find a random walker on a node of degree $k$ in a connected and
undirected network with $M_{0}$ links is given by \cite{Noh04}
\begin{equation}
p_{k}=\frac{k}{2M_{0}}.
\label{eq:steadyprob}
\end{equation}
Using this relation, the probability $h_{k}(w)$ that $w$
walkers are observed on a node of degree $k$ is presented by
\begin{equation}
h_{k}(w)=\binom{W_{0}}{w}p_{k}^{w}(1-p_{k})^{W_{0}-w},
\label{eq:binom}
\end{equation}
where $W_{0}$ is the total number of walkers in the network. This
leads a natural definition of the node capacity $q_{k}$ of a node
of degree $k$ as
\begin{equation}
q_{k}=\langle w\rangle_{k} + m\sigma_{k},
\label{eq:capacity}
\end{equation}
where $\langle w\rangle_{k}$ and $\sigma_{k}$ are the average
and the standard deviation of the binomial distribution
$h_{k}(w)$, which are given by $\langle w\rangle_{k}=W_{0}p_{k}$
and $\sigma_{k}=\sqrt{W_{0} p_{k}(1-p_{k})}$, respectively, and
$m$ is a real positive parameter which characterizes the node
tolerance to load. Since the overload probability $F_{W_{0}}(k)$
of a node of degree $k$ is the probability of $w$ to exceed
$q_{k}$, $F_{W_{0}}(k)$ is given by summing up the distribution
function Eq.~(\ref{eq:binom}) over $w$ larger than $q_{k}$. Thus,
we have \cite{Kishore11}
\begin{eqnarray}
F_{W_{0}}(k) &=&  \sum_{w=\lfloor q_{k}\rfloor +1}^{W_{0}}
                \binom{W_{0}}{w}p_{k}^{w}(1-p_{k})^{W_{0}-w} \nonumber \\
             &=&I_{k/2M_{0}}(\lfloor q_{k}\rfloor +1,W_{0}-\lfloor q_{k}\rfloor ),
\label{eq:FWk}
\end{eqnarray}
where $I_{p}(a,b)$ is the regularized incomplete beta function
\cite{Abramowitz64} and $\lfloor x\rfloor$ is the greatest
integer not greater than $x$. It is important to pay attention
to the fact that the overload probability is a decreasing
function of degree $k$ \cite{Kishore11}.

Based on the above overload probability, we model the cascade
process of overload failures as follows:

(i) Prepare an initial connected, uncorrelated, and undirected
network $\mathcal{G}_{0}$ with $N_{0}$ nodes and $M_{0}$ links,
in which $W_{0}$ random walkers exist, and determine the
capacity $q_{k}$ of each node according to
Eq.~(\ref{eq:capacity}).

(ii) At each time step $\tau$, assign $W_{\tau}$ random walkers
to the network $\mathcal{G}_{\tau}$ at time $\tau$. The total
load $W_{\tau}$ is given by
\begin{equation}
W_{\tau}=\left( \frac{M_{\tau}}{M_{0}} \right)^{r} W_{0},
\label{eq:Wtau}
\end{equation}
where $M_{\tau}$ is the total number of links in the network
$\mathcal{G}_{\tau}$ and $r$ is a real positive parameter.

(iii) Calculate the overload probability of every node, and
remove nodes from $\mathcal{G}_{\tau}$ with this probability.

(iv) Repeat (ii) and (iii) until no node is removed in the
procedure (iii).

In the procedure (ii), the total load $W_{\tau}$ is reduced in
accordance with the reduction of the network size. In actual
cases of cascading failures, the total load is often reduced to
some extent during a cascade process to prevent the breakdown
of the network function. For instance, when a problem arises in
a power supply system due to a natural disaster, the temporary
restriction of electricity use is sometimes introduced to avoid
a large-scale breakdown of the power grid, as in the case of
the Tohoku earthquake and tsunami in 2011 \cite{Arikawa14}. In
addition, when a company goes bankrupt, a large-scale chain
bankruptcy could be prevented by the reduction of the total
debt (loads) on the transaction network by means of, for
example, a special low-interest lending facilities for
companies having business relationships with the bankrupt
company or the injection of taxpayers' money. The quantity
$W_{\tau}$ given by Eq.~(\ref{eq:Wtau}) represents such a
reduction of the total load during the cascade process. The
exponent $r$ characterizes how quickly the total load decreases
with decreasing the network size, which we call hereafter the
load reduction parameter. Although the initial network
$\mathcal{G}_{0}$ is connected, the network $\mathcal{G}_{\tau}$
at cascade step $\tau$ is not necessarily connected. For an
unconnected network $\mathcal{G}_{\tau}$, we assume that
$W_{\tau}$ random walkers are distributed to components in
proportion to the numbers of links in these components. Namely,
the load allocated to the $\alpha$-th component is given by
\begin{equation}
W_{\tau}^{\alpha}=\frac{M_{\tau}^{\alpha}}{M_{\tau}} W_{\tau},
\label{eq:Wtaualpha}
\end{equation}
where $M_{\tau}^{\alpha}$ is the number of links in the
$\alpha$-th component.

In the procedure (iii), the overload probability during the
cascade process cannot be calculated straightforwardly by
Eq.~(\ref{eq:FWk}). First, the degree $k$ of a node in the
network $\mathcal{G}_{\tau}$ at step $\tau$ is not the same as
its initial degree $k_{0}$. Since the probability to find a
random walker on a node in $\mathcal{G}_{\tau}$ is a function
of the present degree $k$ of the node while the node capacity
is determined by its initial degree $k_{0}$, the overload
probability is presented by Eq.~(\ref{eq:FWk}) with replacing
$q_{k}$ by $q_{k_{0}}$. Moreover, when the network
$\mathcal{G}_{\tau}$ is unconnected, the overload probability
must be calculated in each component, because random walkers in
a component cannot move beyond the component. Therefore, the
overload probability of a node of degree $k$, whose initial
degree is $k_{0}$, in the $\alpha$-th component is given by
\begin{equation}
F_{W_{\tau}^{\alpha}}(k_{0},k)=I_{k/2M_{\tau}^{\alpha}}
\left(\lfloor q_{k_{0}}(W_{0})\rfloor +1, W_{\tau}^{\alpha}-\lfloor q_{k_{0}}(W_{0})\rfloor \right),
\label{eq:Ftaualpha}
\end{equation}
where $M_{\tau}^{\alpha}$ and $W_{\tau}^{\alpha}$ are again the
total number of links and the number of random walkers in the
$\alpha$-th component of $\mathcal{G}_{\tau}$, respectively.

The robustness of a network against cascading overload failures
described above is evaluated by the relative size of the giant
component $S_{\text{f}}\equiv N_{\text{f}}/N_{0}$, where
$N_{\text{f}}$ is the number of nodes in the largest component
of the network $\mathcal{G}_{\text{f}}$ at the final stage of
the cascade process. More specifically, the robustness of the
network is measured by the load reduction parameter
$r_{\text{c}}$ above which the relative size $S_{\text{f}}$
becomes finite. A network providing a smaller $r_{\text{c}}$ can
be regarded to be more robust in the sense that there exists a
giant component even if the total load is slowly reduced in
accordance with the reduction of the network size during the
cascade process.

\section{SIZE OF THE GIANT COMPONENT}
\label{sec:3}
In order to assess the robustness of a network, we calculate
the relative size $S_{\tau}$ of the giant component in the
network $\mathcal{G}_{\tau}$ at cascade step $\tau$. The
calculation of $S_{\tau}$ requires information on the load
$W_{\tau}^{\alpha}$ to obtain the overload probability
$F_{W_{\tau}^{\alpha}}(k_{0},k)$ for each component. It is,
however, difficult to find an analytical expression of
$W_{\tau}^{\alpha}$. Thus, we assume that random walkers can
jump to other components with a small probability, which
enables us to estimate the stationary probability to find a
walker on a node of degree $k$ by
\begin{equation}
p_{k}=\frac{k}{2M_{\tau}},
\label{eq:steadyprob2}
\end{equation}
instead of $p_{k}^{\alpha}=k/2M_{\tau}^{\alpha}$, and the
overload probability by
\begin{equation}
F_{W_\tau}(k_{0},k)=I_{k/2M_{\tau}}\left(\lfloor q_{k_{0}}(W_{0})\rfloor
+1, W_{\tau}-\lfloor q_{k_{0}}(W_{0})\rfloor \right),
\label{eq:Ftau}
\end{equation}
instead of Eq.~(\ref{eq:Ftaualpha}). This simplification does
not largely change the overload probability if the number of
links in a component is large enough. The reason is the
following. The probability that $w$ walkers are found on a
degree-$k$ node in the $\alpha$-th component containing
$W_{\tau}^{\alpha}$ walkers is written as
\begin{equation}
h(w;W_{\tau}^{\alpha},p_{k}^{\alpha})=\binom{W_{\tau}^{\alpha}}{w}
\left(p_{k}^{\alpha}\right)^{w}\left(1-p_{k}^{\alpha}\right)^{W_{\tau}^{\alpha}-w},
\label{eq:binom_taualpha}
\end{equation}
if random walkers are confined in the component. The average
and the standard deviation of this binomial distribution
function of $w$ are $\langle
w\rangle_{k}=W_{\tau}^{\alpha}p_{k}^{\alpha}=
W_{\tau}k/2M_{\tau}$ and $\sigma_{k}=\sqrt{\langle
w\rangle_{k}(1-k/2M_{\tau}^{\alpha})}$, respectively, where we
used Eq.~(\ref{eq:Wtaualpha}) and the relation
$p_{k}^{\alpha}=k/2M_{\tau}^{\alpha}$. In the case that random
walkers are allowed to jump between components, on the other
hand, the probability to find $w$ walkers on a degree-$k$ node
is given by $h(w;W_{\tau},p_{k})$. The average $\langle
w\rangle_{k}'$ of this distribution function coincides with the
average $\langle w\rangle_{k}$ of
$h(w;W_{\tau}^{\alpha},p_{k}^{\alpha})$. Although the standard
deviation $\sigma_{k}'=\sqrt{\langle
w\rangle_{k}(1-k/2M_{\tau})}$ of $h(w;W_{\tau},p_{k})$ is
larger than $\sigma_{k}$ of
$h(w;W_{\tau}^{\alpha},p_{k}^{\alpha})$, $\sigma_{k}'$ is not
very different from $\sigma_{k}$ if $M_{\tau}^{\alpha}$ (thus
$M_{\tau}$) is large enough. Therefore, both distribution
functions $h(w;W_{\tau}^{\alpha},p_{k}^{\alpha})$ and
$h(w;W_{\tau},p_{k})$ with the same average and similar widths
are close to each other. Because of this similarity,
$F_{W_{\tau}^{\alpha}}(k_{0},k)=\sum_{w\ge \lfloor
q_{k_{0}}\rfloor +1}h(w;W_{\tau}^{\alpha},p_{k}^{\alpha})$
that leads Eq.~(\ref{eq:Ftaualpha}) can be approximated by
$F_{W_{\tau}}(k_{0},k)=\sum_{w\ge \lfloor q_{k_{0}}\rfloor
+1}h(w;W_{\tau},p_{k})$ giving Eq.~(\ref{eq:Ftau}).

We calculate the relative size $S_{\tau}$ of the giant
component by using the generating function method
\cite{Newman01}. To this end, here we slightly modify the rule
of the procedure (ii) described in Sec.~II so that overloaded
nodes are not removed but left in the system as zero-degree
nodes for which random walkers never visit. This does not
influence any results in this work, except that the total
number of nodes in $\mathcal{G}_{\tau}$ remains constant at
$N_{0}$, which makes the theoretical treatment easier. In spite
of this modification, we will continue to use the expression
``remove a node" for simplicity, but this actually means
``remove all links from a node". The generating function method
then enables us to calculate $S_{\tau}$ if the network
$\mathcal{G}_{\tau}$ is uncorrelated and the degree
distribution function $P_{\tau}(k)$ of $\mathcal{G}_{\tau}$ is
given. In order to estimate $P_{\tau}(k)$, we introduce the
probability $\Pi_{\tau}(k_{0},k)$ that a randomly chosen node
has the degree $k$ and the initial degree $k_{0}$. Obviously,
the probability $\Pi_{\tau}(k_{0},k)$ is related to the degree
distribution $P_{\tau}(k)$ through
\begin{equation}
P_{\tau}(k)=\sum_{k_{0}\ge k}\Pi_{\tau}(k_{0},k).
\label{eq:ptauk}
\end{equation}
As a special case, we have, at $\tau=0$,
\begin{equation}
\Pi_{0}(k_{0},k)=P_{0}(k)\delta_{k_{0}k},
\label{eq:Pi0}
\end{equation}
because $k$ is always equal to $k_{0}$ in the initial network
$\mathcal{G}_{0}$. To obtain the probability
$\Pi_{\tau}(k_{0},k)$, we further introduce the probability
$\phi_{\tau}(k)$ that the overload failure occurs on a node
connected to a node of degree $k$ in the network
$\mathcal{G}_{\tau}$. Considering that
$\Pi_{\tau}(k_{0},k)/P_{\tau}(k)$ represents the probability of
a degree-$k$ node chosen randomly from $\mathcal{G}_{\tau}$ to
have the initial degree $k_{0}$, the probability $\phi_{\tau}(k)$
is expressed as
\begin{equation}
\phi_{\tau}(k) =\sum_{k_{0}}\sum_{k'=1}^{k_{0}}P_{\tau}(k'|k)
\frac{\Pi_{\tau}(k_{0},k')}{P_{\tau}(k')}F_{W_{\tau}}(k_{0},k'),
\label{eq:phi1}
\end{equation}
where $P_{\tau}(k'|k)$ is the conditional probability that a
node of degree $k$ is connected to a node of degree $k'$. Since
the network $\mathcal{G}_{0}$ has no degree correlations,
$\mathcal{G}_{\tau}$ generated by removing nodes randomly from
$\mathcal{G}_{0}$ with a probability depending only on the
degree is also uncorrelated \cite{Srivastava12}. Therefore, the
conditional probability is presented by
$P_{\tau}(k'|k)=k'P_{\tau}(k')/\langle k\rangle_{\tau}$, where
$\langle k\rangle_{\tau}$ is the average degree of
$\mathcal{G}_{\tau}$, and $\phi_{\tau}$ is written as
\begin{equation}
\phi_{\tau} =\sum_{k_{0}}\sum_{k'=1}^{k_{0}}\frac{k'\Pi_{\tau}(k_{0},k')}{\langle k\rangle_{\tau}}F_{W_{\tau}}(k_{0},k'),
\label{eq:phi2}
\end{equation}
which is independent of $k$. The probability $\Pi_{\tau}(k_{0},k)$
is equal to the probability that a randomly chosen node from
$\mathcal{G}_{\tau-1}$ has the initial degree $k_{0}$ and the degree
of this node becomes $k$ during the cascade from $\mathcal{G}_{\tau-1}$
to $\mathcal{G}_{\tau}$. Thus, we can set up the master equation
for $\Pi_{\tau}(k_{0},k)$ as
\begin{widetext}
\begin{equation}
\Pi_{\tau}(k_{0},k)=\sum_{k'\ge k}\Pi_{\tau-1}(k_{0},k')
\Biggl\{\binom{k'}{k}\phi_{\tau-1}^{k'-k}(1-\phi_{\tau-1})^{k}
[1-F_{W_{\tau-1}}(k_{0},k')]+\delta_{k0}F_{W_{\tau-1}}(k_{0},k')\Biggr\}.
\label{eq:Pikk}
\end{equation}
\end{widetext}
The right-hand side of this equation represents the probability
that a degree-$k'$ node in $\mathcal{G}_{\tau-1}$ becomes a
node of degree $k$. The first term describes the situation that
the degree-$k'$ node survives and $k'-k$ nodes adjacent to this
node are removed by overload failures. The second term stands
for the case that the degree-$k'$ node itself experiences an
overload failure and becomes a zero-degree node. Solving
numerically Eq.~(\ref{eq:Pikk}) with the aid of
Eq.~(\ref{eq:phi2}), we can obtain the time evolution of the
the probability $\Pi_{\tau}(k_{0},k)$ and the degree
distribution $P_{\tau}(k)$ by Eq.~(\ref{eq:ptauk}). The
relative size $S_{\tau}$ of the giant component at cascade step
$\tau$ is calculated by \cite{Newman01}
\begin{equation}
S_{\tau}=1-\sum_{k}P_{\tau}(k)u^{k},
\label{eq:Stau}
\end{equation}
where $u$ is the smallest non-negative solution of the transcendental
equation,
\begin{equation}
u=G_{1}^{(\tau)}(u),
\label{eq:u}
\end{equation}
and $G_{1}^{(\tau)}(x)$ is the generating function of the
remaining degree distribution, which is defined by
\begin{equation}
G_{1}^{(\tau)}(x)=\frac{1}{\langle k\rangle_{\tau}}\sum_{k}(k+1)P_{\tau}(k+1)x^{k}.
\label{eq:G1}
\end{equation}

According to the procedure (iv) stated in Sec.~\ref{sec:2}, the
cascade process must be terminated when no node experiences
overload failures. In an actual calculation, we stop the
cascade process  at the step $\tau$ satisfying the condition
\begin{equation}
\sum_{k,k_{0}}F_{W_{\tau}}(k_{0},k)\Pi_{\tau}(k_{0},k)<\frac{1}{N_{0}}.
\label{eq:eqcond}
\end{equation}
This condition implies that the expectation number of overloaded
nodes becomes less than unity.

\begin{figure}[tttt]
\begin{center}
\includegraphics[width=0.48\textwidth]{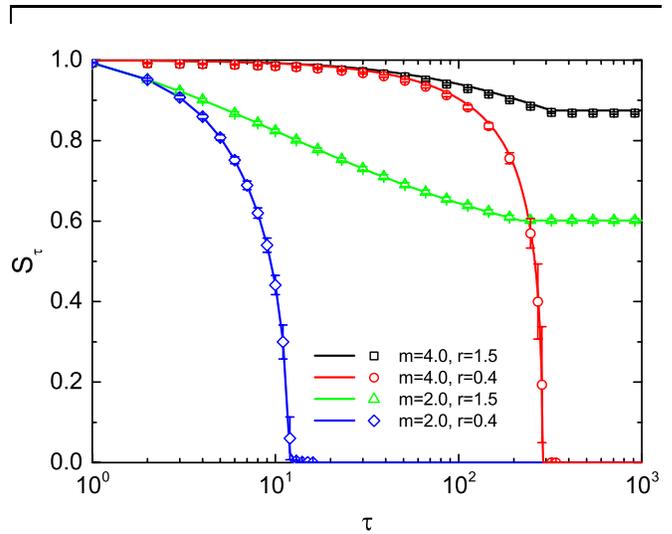}
\caption{
(Color online) Relative size $S_{\tau}$ of the giant component
as a function of the cascade step $\tau$. The initial network
is an Erd\H{o}s-R\'enyi random graph with $5,000$ nodes and
$12,500$ links, which implies that the initial average degree
is $\langle k\rangle_{0}=5.0$. The symbols indicate the results
obtained by numerical simulations following the cascade process
from (i) to (iv) described in Sec.~\ref{sec:2} and averaging
over $50$ network realizations. Error bars represent one standard
deviation from the mean values. The lines show the
results calculated by the analytical method explained in
Sec.~\ref{sec:3}. The total number of walkers at the initial
cascade step is $W_{0}=25,000$. The values of the node tolerance
parameter $m$ and the load reduction parameter $r$ are
displayed in the figure.
}
\label{fig:1}
\end{center}
\end{figure}
It should be noted that the above formalism is based on the
approximation that random walkers are allowed to jump over
components with a small probability. In order to evaluate the
accuracy of this approximation, we compare the time evolutions
of $S_{\tau}$ for an Erd\H{o}s-R\'{e}nyi random graph
calculated by both the method explained in this Section and
numerical simulations following the cascade process from (i) to
(iv) described in Sec.~\ref{sec:2}. The results shown in
Fig.~\ref{fig:1} indicate that $S_{\tau}$ calculated by
Eq.~(\ref{eq:Stau}) agrees quite well with the simulation
result, which reflects the high accuracy of the approximation.

\section{RESULTS}
\label{sec:4}
\begin{figure}[tttt]
\begin{center}
\includegraphics[width=0.48\textwidth]{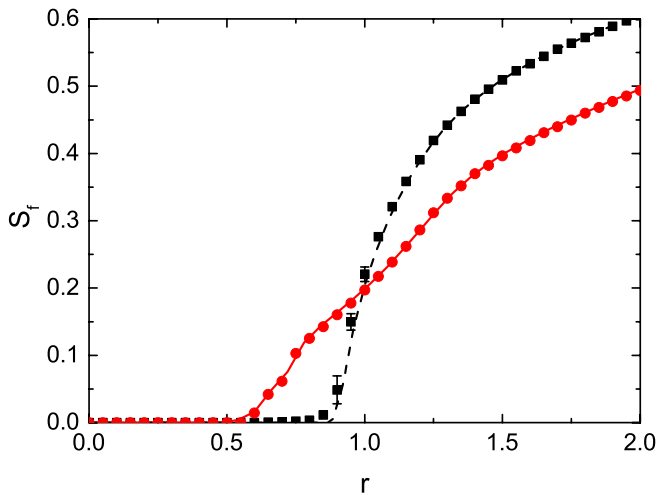}
\caption{
(Color online) Relative size $S_{\text{f}}$ of the giant
component at the final stage of the cascade process as a
function of the load reduction parameter $r$, for an ER random
graph (black dashed line and filled squares) and an SF network
(red solid line and filled circles). The SF network has the
degree distribution given by Eq.~(\ref{eq:pk}) with
$\gamma=2.5$. The lines represent the results calculated by the
analytical method explained in Sec.~\ref{sec:3}. The symbols
indicate the simulation results obtained by following
numerically the cascade process from (i) to (iv) described in
Sec.~\ref{sec:2} and averaging over $50$ samples. Only error
bars larger than the symbol size are shown. In this
calculation, we set $N_{0}=10^{4}$, $M_{0}=2\times 10^{4}$,
$m=2.0$, and $W_{0}=2M_{0}$.
}
\label{fig:2}
\end{center}
\end{figure}
We assess the robustness of networks against cascading overload
failures by computing the relative size $S_{\text{f}}$ of the
giant component at the final stage of the cascade process as a
function of the load reduction parameter $r$. In this work, we
examined two types of initial networks, namely, homogeneous
Erd\H{o}s-R\'enyi (ER) random graphs and scale-free (SF)
networks with inhomogeneous degree distributions. An SF network
is generated by the configuration model \cite{Molloy95} with
the degree distribution given by
\begin{equation}
P_{0}(k)=
\begin{cases}
\displaystyle
\frac{C}{k^{\gamma}+d^{\gamma}} & \text{for } k\ge k_{\text{min}},\\
0 & \text{for } k< k_{\text{min}},
\end{cases}
\label{eq:pk}
\end{equation}
where $d$ is a real positive parameter and $\gamma$ is the
exponent describing the asymptotic power-law form of the degree
distribution, i.e., $P_{0}(k)\propto k^{-\gamma}$ for $k\gg d$.
In Eq.~(\ref{eq:pk}), $k_{\text{min}}$ is the minimum degree,
and $C$ is the normalization constant. The parameters $d$ and
$k_{\text{min}}$ can control the average degree $\langle
k\rangle_{0}$ for a fixed value of $\gamma$. Throughout this paper
the minimum degree is fixed at $k_{\text{min}}=2$. Figure
\ref{fig:2} shows the $r$ dependence of $S_{\text{f}}$ for an
ER random graph (black dashed line and filled squares) and for
an SF network (red solid line and filled circles). Both networks
have $N_{0}=10^{4}$ nodes and $M_{0}=2\times 10^{4}$ links,
which implies that the initial average degree is $\langle
k\rangle_{0}=4.0$. The initial total load is chosen as
$W_{0}=2M_{0}$ and the node capacity $q_{k}$ is determined by
Eq.~(\ref{eq:capacity}) with $m=2.0$. For the SF network, the
exponent $\gamma$ is set as $\gamma=2.5$ and $d$ is tuned to
satisfy $\langle k\rangle_{0}=4.0$. The solid and dashed lines
in Fig.~\ref{fig:2} indicate the results calculated by the
analytical method explained in Sec.~\ref{sec:3}, whereas the
symbols represent $S_{\text{f}}$ obtained by simulating
numerically the cascade process from (i) to (iv) described in
Sec.~\ref{sec:2}. For both types of networks, the analytical
results agree quite well with the numerical ones. As shown in
Fig.~\ref{fig:2}, there exists a value $r_{\text{c}}(N_{0})$
below which $S_{\text{f}}=0$ and above which $S_{\text{f}}>0$.
This implies that a global cascade of overload failures occurs
if the total load $W_{\tau}$ is reduced during the cascade more
slowly than the reduction scheme Eq.~(\ref{eq:Wtau}) with
$r=r_{\text{c}}(N_{0})$ while a finite fraction of nodes
survives the cascade if the reduction of $W_{\tau}$ is faster.
The fact that the value of $r_{\text{c}}(N_{0})$ for the SF
network is smaller than that for the ER random graph suggests
that SF networks are more robust to cascading overload failures
than ER random graphs. This tendency is opposite to what
was predicted by previous works \cite{Motter02,Holme02b,
Crucitti04,Wu07,Bao09,Xia10,Dou10}.

\begin{figure}[tttt]
\begin{center}
\includegraphics[width=0.48\textwidth]{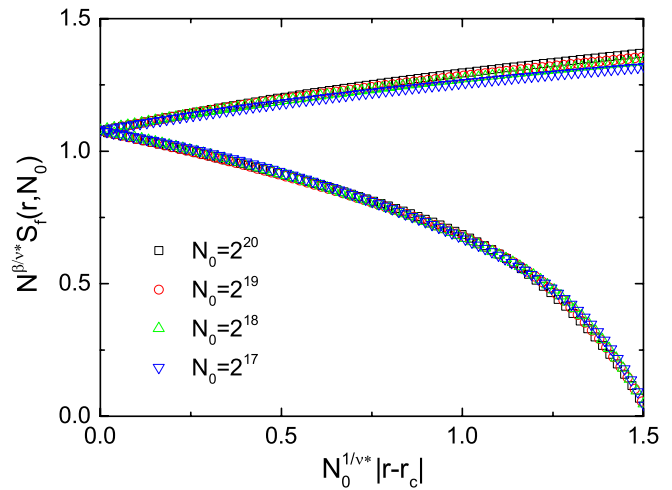}
\caption{
(Color online) Scaling plot of $S_{\text{f}}(r,N_{0})$ for ER
random graphs of several sizes ($N_{0}=2^{17},2^{18}, 2^{19}$,
and $2^{20}$). The average degree of these networks is set to
be $\langle k\rangle_{0}=4.0$. The calculations have been done
under the condition of $W_{0}=N_{0}\langle k\rangle_{0}$ and
$m=2.0$. In this plot, we choose $r_{\text{c}}=1.84$,
$\beta=1.27$, and $1/\nu^{*}_{r}=19.45$. Two branches
correspond to the percolating and non-percolating phases.
}
\label{fig:3}
\end{center}
\end{figure}
The above conclusion is, however, obtained for finite-size
networks. It is crucial to study the robustness of infinitely
large SF networks against cascading overload failures. If the
change of $S_{\text{f}}$ at $r=r_{\text{c}}(N_{0})$ provides a
critical transition when $N_{0}$ goes to infinity, we can use
the finite-size scaling analysis to calculate $r_{\text{c}}$ in
the thermodynamic limit. According to the finite-size scaling
theory, the relative size of the giant component in the network
$\mathcal{G}_{\text{f}}$ at the final stage of the cascade
process starting from the initial network $\mathcal{G}_{0}$ of
size $N_{0}$ is expressed as
\begin{equation}
S_{\text{f}}(r,N_{0})=N_{0}^{-\beta/\nu^{*}}\Phi\left(N_{0}^{1/\nu^{*}}
| r-r_{\text{c}}|\right),
\label{eq:fss}
\end{equation}
where the correlation volume exponent $\nu^{*}$ characterizes
the divergence of the number of nodes $N_{\xi}$ within the
correlation volume as $N_{\xi}\propto |r-r_{\text{c}}|^{-\nu^{*}}$
in the infinite system, $\beta$ is the critical exponent for
the approach to zero of $S_{\text{f}}(r,\infty)$,
$r_{\text{c}}$ is the critical load reduction parameter in the
thermodynamic limit, and $\Phi(x)$ is a scaling function.
Therefore, if the suitable values of the parameters $\nu^{*}$,
$\beta$, and $r_{\text{c}}$ are selected, the quantity
$N_{0}^{\beta/\nu^{*}}S_{\text{f}}(r,N_{0})$ as a function of
$N_{0}^{1/\nu^{*}}| r-r_{\text{c}}|$ collapses onto a single
curve for various values of $r$ and $N_{0}$. Figure \ref{fig:3}
shows such a plot for ER random graphs of different sizes with
the use of the $r$ dependence of $S_{\text{f}}(r,N_{0})$
calculated by the method explained in Sec.~\ref{sec:3} and the
best-fit values of $\nu^{*}$, $\beta$, and $r_{\text{c}}$. The
fact that all data collapse onto a single curve implies that
the transition at $r=r_{\text{c}}$ can be considered as a
critical phenomenon. Similar scaling behaviors have been
confirmed for SF networks.

\begin{figure}[tttt]
\begin{center}
\includegraphics[width=0.48\textwidth]{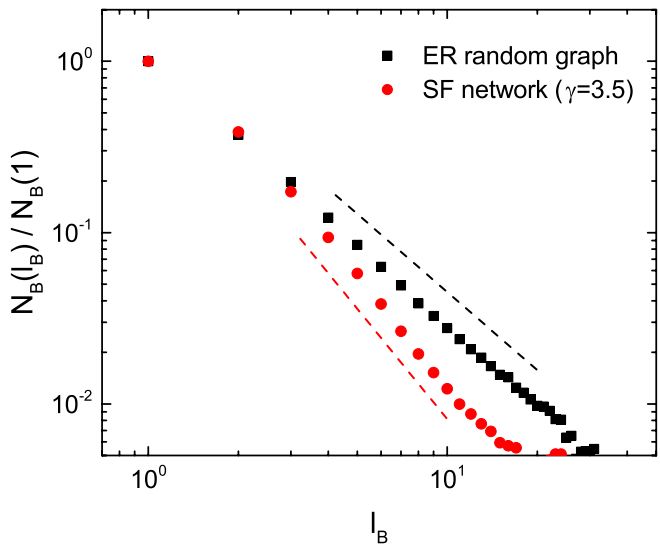}
\caption{
(Color online) $N_{\text{B}}(l_{\text{B}})$ for the giant
components in $\mathcal{G}_{\text{f}}$'s obtained numerically
by simulating the cascade process with $r=r_{\text{c}}(N_{0})$
starting from an ER random graph (black filled squares) and an
SF network with $\gamma=3.5$ (red filled circles). The number
of nodes and the average degree of both of the initial networks
are $N_{0}=10^4$ and $\langle k\rangle_{0}=4.0$, respectively.
The calculations have been done under the condition of
$W_{0}=N_{0}\langle k\rangle_{0}$ and $m=2.0$. The results are
averaged over $100$ samples. The longitudinal axis is rescaled
by $N_{\text{B}}(l_{\text{B}})$ at $l_{\text{B}}=1$. The dashed
lines are merely guides to the eye.
}
\label{fig:4}
\end{center}
\end{figure}
The criticality of the network $\mathcal{G}_{\text{f}}$ at
$r=r_{\text{c}}$ has also been confirmed by the fractal property of
the giant component in $\mathcal{G}_{\text{f}}$. The fractality of
complex networks is widely investigated by the box covering algorithm
\cite{Song05,Song06,Goh06,Kim07,Kawasaki10,Furuya11,Sun14}. If
the minimum number $N_{\text{B}}(l_{\text{B}})$ of subgraphs of
radius $l_{\text{B}}$ required to cover a given connected
network satisfies the relation
\begin{equation}
N_{\text{B}}(l_{\text{B}})\propto l_{\text{B}}^{-d_{\text{B}}},
\label{eq:fractal}
\end{equation}
the network is considered to be fractal with the fractal
dimension $d_{\text{B}}$ \cite{Song05}. We calculated, by using
the compact-box-burning algorithm \cite{Song07},
$N_{\text{B}}(l_{\text{B}})$ for giant components included in
networks $\mathcal{G}_{\text{f}}$ at the final stage of the
cascade process with $r=r_{\text{c}}(N_{0})$ starting from both
an ER random graph and SF network. The giant components are
obtained numerically by simulating the cascade process
described in Sec.~\ref{sec:2}. The results shown in
Fig.~\ref{fig:4} indicate that the structures of these giant
components exhibit the fractal nature, which supports the
criticality of $\mathcal{G}_{\text{f}}$ at $r=r_{\text{c}}$.
The fractal dimension for the ER random graph is
$d_{\text{B}}=1.54\pm 0.01$, while $d_{\text{B}}=2.13\pm 0.02$
for the SF network with $\gamma=3.5$. These values of
$d_{\text{B}}$ are different from those of giant components at
the critical point of the conventional percolation with random
node removals, which are $d_{\text{B}}=2$ for ER random graphs
(or SF networks with $\gamma\ge 4$) and
$d_{\text{B}}=(\gamma-2)/(\gamma-3)$ for SF networks with
$3<\gamma<4$ \cite{Cohen04}. Such a discrepancy is, of course,
due to the difference in ways of node removals. In the cascade
process, nodes that will be removed at the cascade step $\tau$
depends strongly on nodes removed at $\tau-1$, as in the case
of a fire spread for which a portion remaining unburned has a
different structure from that of a survival from random
removals.

\begin{figure}[tttt]
\begin{center}
\includegraphics[width=0.48\textwidth]{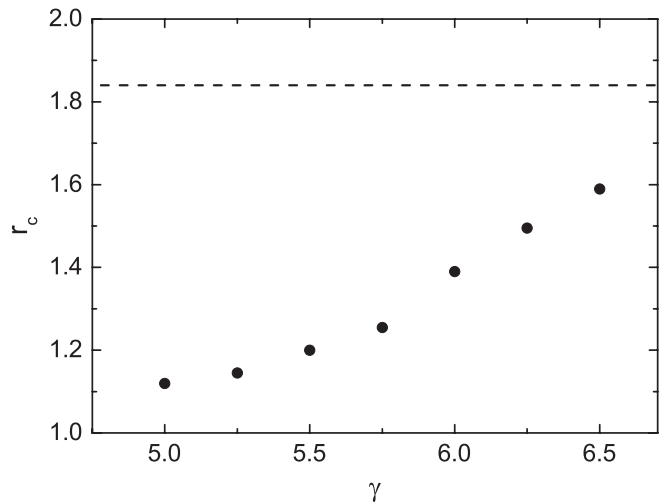}
\caption{
Critical load reduction parameter $r_{\text{c}}$ as a function
of the exponent $\gamma$ in the thermodynamic limit. In this
calculation for SF networks, we tune the parameter $d$ in
Eq.~(\ref{eq:pk}) to satisfy $\langle k\rangle_{0}=4.0$ for
various values of $\gamma$. The values of $r_{\text{c}}$ are
estimated through the finite-size scaling analysis for
$S_{\text{f}}(r,N_{0})$ calculated by the method described in
Sec.~\ref{sec:3} under the condition of $W_{0}=\langle
k\rangle_{0}N_{0}$ and $m=2.0$. The horizontal dashed line is
the result for the ER random graph ($\gamma\rightarrow\infty$)
with $\langle k\rangle_{0}=4.0$.
}
\label{fig:5}
\end{center}
\end{figure}
The critical load reduction parameter $r_{\text{c}}$ calculated
by the finite-size scaling method is plotted in
Fig.~\ref{fig:5} as a function of the exponent $\gamma$ of SF
networks. In this figure, $r_{\text{c}}$ for the ER random
graph is also indicated by the horizontal dashed line. The
results only for $\gamma\ge 5$ are presented here, because for
a small value of $\gamma$ the computation of
$\Pi_{\tau}(k_{0},k)$ requires a long CPU time due to the
increase of the maximum degree $k_{\text{max}}$ of the initial
network $\mathcal{G}_{0}$ associated with a decrease of $\gamma$
as is given by $k_{\text{max}}\propto N_{0}^{1/\gamma}$. We see
from Fig.~\ref{fig:5} that $r_{\text{c}}$ is an increasing
function of $\gamma$. This implies that the enhancement of the
SF property by decreasing $\gamma$ makes networks robust to
cascading overload failures even in the thermodynamic limit. It
should be emphasized that $r_{\text{c}}$ is always positive for
any $\gamma$. This is because in any network a cascade of
overload failures under $r=0$ never stops until the network
collapses completely. Our result contrary to previous
predictions \cite{Motter02,Holme02b,Crucitti04,Wu07,Bao09,Xia10,
Dou10} comes from the fact that the overload probability is a
decreasing function of degree $k$. At the first step of the
cascade process, nodes with small degrees are more likely to be
removed according to the overload probability $F_{W_{0}}(k)$. A
low-degree node tends to be connected to a node with large
degree in an uncorrelated SF network. The overload probability
of a high-degree node adjacent to the low-degree node which was
removed at the first step becomes smaller at the second cascade
step than its initial overload probability. Therefore, nodes
with relatively small degrees are again preferentially removed
also at the second step, and so on. It is obvious that SF
networks are robust against preferential removals of low-degree
nodes \cite{Moreira09}.

Our model differs from previous models of cascading overload
failures in two points. One is that overload failures in our
model occur when fluctuating loads exceed the capacities
predetermined for nodes, while failures are caused by
non-fluctuating loads (or average values of fluctuating loads)
exceeding the node capacities in previous models
\cite{Motter02,Holme02b,Crucitti04,Wu07,Bao09,Xia10,Dou10}. The
other difference is in the dynamics of loads on a network. In
many of previous works \cite{Motter02,Holme02b,Crucitti04,Wu07,
Bao09,Xia10,Dou10}, the load is defined by flow passing through
the shortest path between a pair of nodes. This type of loads
describes, for example, a packet flow in the Internet and a
traffic flow in a road system. On the other hand, loads in our
model move randomly on a network, as in the case of a flow of
debt in a corporate transaction network. It is important to
clarify whether the robustness of SF networks is caused by the
fluctuation of load or by its random walk behavior.
\begin{figure}[tttt]
\begin{center}
\includegraphics[width=0.48\textwidth]{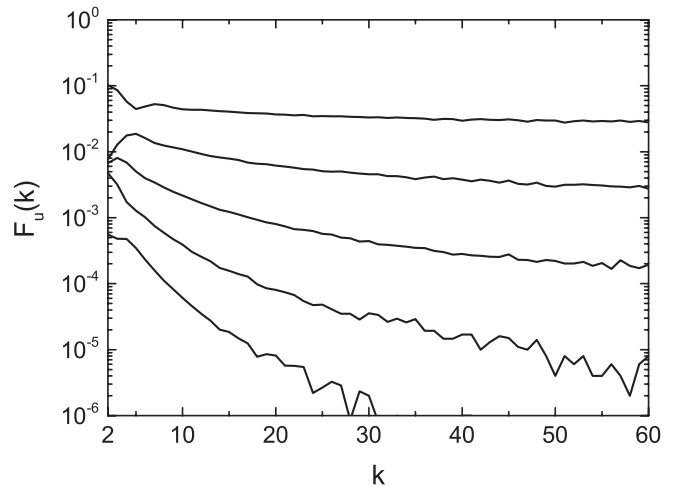}
\caption{
Degree dependence of the overload probability $F_{u}(k)$ for
several values of the node tolerance parameter $m$. The lines
from top to bottom represent the results for $m=2.0$, $3.0$,
$4.0$, $5.0$, and $6.0$, respectively. The size of the set
$V_{u}$ is fixed at $u=500$. The load distribution
$\tilde{h}_{k}(w)$ required to calculate $F_{u}(k)$ is
numerically evaluated by computing the partial betweenness
centrality for each of $10^{4}$ different sets $V_{u}$ selected
from each of $30$ SF networks with $10^{4}$ nodes. The degree
distribution of these SF networks is given by Eq.~(\ref{eq:pk})
with $\gamma=2.5$ and $d=0.152$ providing $\langle
k\rangle=4.0$.
}
\label{fig:6}
\end{center}
\end{figure}

In order to identify the origin of the robustness, namely, the
origin of $F_{W_{0}}(k)$ being a decreasing function of degree
$k$, we calculate the overload probability for fluctuating
loads imposed by a shortest-path flow. Fluctuating loads
carried by flow along the shortest paths have been argued in
Ref.~\cite{Menezes04} to explain the relation between the
average flux and the fluctuations. According to this work, we
newly define the load $w_{i}$ of the node $i$ by
\begin{equation}
w_{i}=\sum_{(j,j')\in V_{u} \atop (j \neq j')}\frac{\sigma_{jj'}(i)}{\sigma_{jj'}},
\label{eq:pbetweenness}
\end{equation}
where $V_{u}$ is a set of $u$ node pairs that are randomly
selected from $(N_{0}-1)(N_{0}-2)/2$ node pairs in
$\mathcal{G}_{0}$ excluding the node $i$, $\sigma_{jj'}$ is the
total number of shortest paths between the pair $(j,j')$ in
$V_{u}$, and $\sigma_{jj'}(i)$ is the number of those paths
that pass through $i$. Since the quantity represented by the
right-hand side of Eq.~(\ref{eq:pbetweenness}) is equivalent to
the betweenness centrality $b_{i}$ if $V_{u}$ is chosen as the
set of all node pairs, namely, $u=(N_{0}-1)(N_{0}-2)/2$, we
call the above quantity $w_{i}$ the \textit{partial betweenness
centrality}. The partial betweenness centrality of a node
depends on which node pairs are selected for the set $V_{u}$,
and thus the load $w_{i}$ fluctuates in accordance with the
choice of $V_{u}$. For a large number of different sets $V_{u}$
with $1\ll u\ll (N_{0}-1)(N_{0}-2)/2$, the distribution
$\tilde{h}_{i}(w)$ of the load can be calculated for each node.
As the conventional betweenness centrality $b_{i}$ is strongly
correlated to the degree $k_{i}$ in an uncorrelated network
\cite{Goh01,Barthelemy04}, the partial betweenness centrality
of a node is also expected to have a correlation with the
degree of the node. In fact, we have confirmed numerically that
two distribution functions $\tilde{h}_{i}(w)$ and
$\tilde{h}_{j}(w)$ of load on the nodes $i$ and $j$ which have
the same degree are close to each other. Therefore, we can
define the degree-dependent load distribution
$\tilde{h}_{k}(w)$ for the partial betweenness centrality
model, which corresponds to the load distribution $h_{k}(w)$
given by Eq.~(\ref{eq:binom}) for the random walker model. The
node capacity $q_{k}$ is also defined by
Eq.~(\ref{eq:capacity}) with the average load $\langle
w\rangle_{k}$ and the standard deviation $\sigma_{k}$
calculated by the distribution function $\tilde{h}_{k}(w)$.
Finally, as in the case of the random walker model, the
overload probability is defined as the probability that the
load of a node of degree $k$ exceeds its capacity $q_{k}$,
which is calculated by
$\int_{q_{k}}^{\infty}\tilde{h}_{k}(w)\,dw$. Since
$\tilde{h}_{k}(w)$ depends on $u$, we denote this overload
probability by $F_{u}(k)$ instead of $F_{W_{0}}(k)$.

Figure \ref{fig:6} shows the degree dependence of $F_{u}(k)$
for several values of the node tolerance parameter $m$. We see
that the overload probability $F_{u}(k)$ is a decreasing
function of $k$ for any $m$. This implies that SF networks are
robust also against cascading overload failures induced by
fluctuating shortest-path flow. It is not surprising that the
degree dependences of $F_{W_{0}}(k)$ and $F_{u}(k)$ show a
similar tendency, because it has been reported that the
couplings between the fluctuations and the average of the
number of random walkers and the partial betweenness centrality
on individual nodes obeys the same scaling law \cite{Menezes04}.
We can also consider a situation that cascading failures are
caused by \textit{non-fluctuating} loads by random walkers. In
this case, an overload failure occurs when the average number
of walkers $\langle w\rangle_{k}=W_{0}k/2M_{0}$ on a node of
degree $k$ exceeds the node capacity depending on $\langle
w\rangle_{k}$, where $W_{0}$ is the total number of walkers and
$M_{0}$ is the number of links in the initial network. Since
$\langle w\rangle_{k}$ is proportional to $k$ and the degree of
a node correlates closely with the betweenness centrality of
the node \cite{Goh01,Barthelemy04}, the property of cascading
failures induced by non-fluctuating random walking loads is
essentially the same as that by loads of the betweenness
centrality \cite{Motter02}. Therefore, SF networks are fragile
to such cascading failures. All the above arguments can be
summarized as shown in Table \ref{table:1}. From this Table, we
can conclude that the robustness of SF networks in our model is
a consequence of the property that failures are caused by
extreme values of fluctuating loads, but not concerned with the
random walk behavior of loads.
\begin{table}[ttt]
  \caption{\label{tab:table1}Robustness/fragility of SF networks against cascading
overload failures induced by several types of loads. The fragility for
shortest path flow without fluctuation is a consequence of the previous
work \cite{Motter02}.}
 \begin{ruledtabular}
  \begin{tabular}{lcc}
    Load flow     & Without fluctuation & With fluctuation \\ \hline
    Shortest path & Fragile & Robust \\
    Random walk   & Fragile & Robust \\
  \end{tabular}
 \end{ruledtabular}
\label{table:1}
\end{table}

\section{CONCLUSIONS}
\label{sec:5}
We have studied the robustness of scale-free (SF) networks
against cascading overload failures induced by extreme values
of fluctuating loads that exceed the node capacities. In our
model, temporally fluctuating loads are treated as random
walkers on a network, for which the stationary overload
probability has been studied by Kishore \textit{et al.}~\cite{Kishore11}.
At the first stage of the cascade, nodes are removed from the
initial network with this overload probability, and the
redistribution of loads in the damaged network causes
iteratively subsequent failures according to the updated
overload probability until no node is expected to be removed.
During the cascade process, the total load is reduced in
response to the decrease of the number of links in the network
under the cascade. How quickly the total load is reduced is
characterized by the load reduction parameter $r$. The
robustness of a network is measured by the critical load
reduction parameter $r_{\text{c}}$ above which the relative
size $S_{\text{f}}$ of the giant component at the final cascade
stage is finite. We present a formulation to calculate
$S_{\text{f}}$ by using the master equation for the probability
$\Pi_{\tau}(k_{0},k)$ of a node in the network $\mathcal{G}_{\tau}$
at cascade step $\tau$ to have the present degree $k$ and the
initial degree $k_{0}$ and by applying the generating function
method. Our results for SF networks with degree distribution
$P_{0}(k)\sim k^{-\gamma}$ at large $k$ show that
$r_{\text{c}}$ increases with the exponent $\gamma$, which
implies that SF networks are robust against cascading overload
failures in our model as opposed to previous works
\cite{Motter02,Holme02b,Crucitti04,Wu07,Bao09,Xia10,Dou10}. The
robustness of SF networks is explained by the property of the
overload probability of being a decreasing function of the
degree, which is a consequence of the load fluctuations but not
concerned with the random walk behavior of loads.

In our model, the total load does not fluctuate throughout the
cascade process though the local load on a node fluctuates. If
the total load also fluctuates temporally and the time scale of
the fluctuation is faster than that of the cascade process, the
overload probability must be different from Eq.~(\ref{eq:FWk})
or (\ref{eq:Ftau}). However, we can expect that the overload
probability remains to be a decreasing function of the degree
if the magnitude of the total load fluctuation $\Delta W$ is
much smaller than the average total load $\langle W\rangle$,
which guarantees the robustness of SF networks to cascading
overload failures. This is because the standard deviation
$\sigma_{k}$ of the fluctuating load $w$ on a node of degree
$k$ is proportional to the square root of the average load
$\langle w\rangle_{k}$ if $\Delta W \ll \langle W\rangle$, as
in the case of the present work. It has, however, been reported
that $\sigma_{k}$ becomes proportional to $\langle
w\rangle_{k}$ when $\Delta W/ \langle W\rangle$ approaches to
$1$ \cite{Menezes04}. The change in the property of the load
fluctuations may alter drastically the degree dependence of the
overload probability and hence the robustness of networks. It
is thus important to study how robust SF networks are against
cascading failures under large fluctuations of the total load.
Furthermore, although we present the formulation to calculate
the relative size of the giant component $S_{\text{f}}$ in this
work, an analytical expression for the critical load reduction
parameter $r_{\text{c}}$ is not found. In addition to solving
this problem, the identification of the universality class of
the present model is also a subject for future work.

\begin{acknowledgements}
This work was supported by a Grant-in-Aid for Scientific
Research (Nos.~25390113 and 14J01323) from the Japan Society
for the Promotion of Science. Numerical calculations in this
work were performed in part on the facilities of the
Supercomputer Center, Institute for Solid State Physics,
University of Tokyo.
\end{acknowledgements}

\end{document}